\documentclass[preprint,12pt,authoryear]{elsarticle}

\usepackage{amssymb}
\usepackage{amsmath, booktabs}
\usepackage{hyperref}
\usepackage[ruled,vlined]{algorithm2e} 
\begin{document}
\begin{frontmatter}

\title{NTLRAG: Narrative Topic Labels derived with Retrieval Augmented Generation}
\author[inst1,inst2]{Lisa Grobelscheg}
\corref{cor1}
\ead{lisa.grobelscheg@s.wu.ac.at, lisa.grobelscheg@campus02.at}

\author[inst1]{Ema Kahr}
\ead{ema.kahr@wu.ac.at}
\author[inst1]{Mark Strembeck}
\ead{mark.strembeck@wu.ac.at}
\affiliation[inst1]{organization={Institute for Complex Networks, Vienna University of Economics and Business (WU Vienna)},
            addressline={Welthandelsplatz 1},
            city={Vienna},
            postcode={1020}, 
            country={Austria}}
\affiliation[inst2]{organization={CAMPUS 02, University of Applied Sciences},
            addressline={Körblergasse 126},
            postcode={8010},
            city={Graz},
            country={Austria}}            
            
\cortext[cor1]{Corresponding author}

\begin{abstract}
Topic modeling has evolved as an important means to identify evident or hidden topics within large collections of text documents. Topic modeling approaches are often used for analyzing and making sense of social media discussions consisting of millions of short text messages. However, assigning meaningful topic labels to document clusters remains challenging, as users are commonly presented with unstructured keyword lists that may not accurately capture the respective core topic. In this paper, we introduce Narrative Topic Labels derived with Retrieval Augmented Generation (NTLRAG), a scalable and extensible framework that generates semantically precise and human-interpretable \emph{narrative topic labels}. Our \emph{narrative topic labels} provide a context-rich, intuitive concept to describe topic model output. In particular, NTLRAG uses retrieval augmented generation (RAG) techniques and considers multiple retrieval strategies as well as chain-of-thought elements to provide high-quality output. NTLRAG can be combined with any standard topic model to generate, validate, and refine narratives which then serve as narrative topic labels. We evaluated NTLRAG with a user study and three real-world datasets consisting of more than 6.7 million social media messages that have been sent by more than 2.7 million users. The user study involved 16 human evaluators who found that our narrative topic labels offer superior interpretability and usability as compared to traditional keyword lists. An implementation of NTLRAG is publicly available for download.
\end{abstract}


\end{frontmatter}

\section{Introduction}

Efforts to automatically summarize large volumes of textual data and derive meaningful information from them resulted in the development of \emph{topic models}. While most topic modeling approaches, such as Latent Dirichlet Allocation (LDA) \citep{lda} are designed for long text documents, some approaches explicitly address the challenges of comparatively short texts, such as social media messages (e.g., Non-Negative Matrix Factorization (NMF), see \citeauthor{nmf}, \citeyear{nmf}). Certain models additionally incorporate metadata, such as a document’s creation date or the document's author, into the modeling process (e.g., Structural Topic Model (STM), see \citeauthor{stm}, \citeyear{stm}). 

While so-called bag-of-words approaches are unaware of the order of words in a document, models based on word embeddings, such as BERTopic or Contextual Topic Models \citep{grootendorst2022bertopicneuraltopicmodeling,bianchi-etal-2021-cross,bianchi-etal-2021-pre} leverage the power of contextual word vectors. \cite{thompson2020topicmodelingcontextualizedword}, suggest that embedding-based models in combination with clustering algorithms are superior compared to traditional LDA models. Some recent approaches also use Large Language Models (LLMs), such as OpenAI's ChatGPT or Google's Gemini, to perform topic modeling tasks \citep{llmtopicmodel}. 

The different flavors of topic modeling approaches are used in a wide variety of application domains. From the impact of Open Innovation \citep{openinnovation}, to trends in finance research \citep{Aziz2021} and the analysis of presidential elections \citep{bidentrump}, researchers have used these methods to investigate different phenomena. A comprehensive overview for the evolution of topic modeling and its applications is provided in \cite{evolutionoftopicmodeling} and \cite{Laureate2023}. 

Thus far, the focus for improving topic models particularly was on achieving higher semantic coherence in document clusters and improvements in cluster composition. However, corresponding approaches do not adequately consider the importance of enabling humans to comprehend the overarching messages conveyed by the documents within such clusters. In our work, we therefore seek to address this gap. 

Typically, topic model output is presented as a set of representative keywords, such as words ranked by their Term-Frequency-Inverse Document Frequency (TF-IDF) scores or raw frequency counts \citep{CHEN2024102322}. An alternative approach involves examining one or more representative document(s) that a specific clustering algorithm identified as a centroid. Therefore, this approach is often employed in conjunction with topic modeling techniques that leverage word embeddings and clustering algorithms \citep{narrativemotifs}. For example, the keyword list of a cluster found in a dataset on a shooting event that we analyzed includes the following terms: 'gambling gambler poker compulsive gambled transactions habits gamble stakes welloff'; the corresponding representative document reads: 'be sure to look into gambling no one makes money playing video poker.' Yet, both the keyword list and the representative document fail to provide an easy-to-interpret, contextualized description of the event. The documents that are included in the corresponding cluster mainly discuss the shooter's gambling habit, a fact that is neither reflected in the keyword list nor in the representative document.

Consistent with prior research \citep{automaticlabelingmultinomaltopicm}, we argue that humans require a more comprehensible description to effectively grasp the core message of a document cluster. Such reader-friendly descriptions are often referred to as \emph{topic labels}.
In this paper, we introduce Narrative Topic Labels with Retrieval Augmented Generation (NTLRAG), an approach that derives narrative topic labels from text corpora as a natural way for people to interpret an event's storyline \citep{weick1995sensemaking} and enhance traditional labels for document clusters.
Prior research \citep{complexis22,snams2022} has conceptualized narratives as: '\textit{a set of topic-wise interconnected messages posted on social media platforms}'. To enhance human interpretation of a narrative, we will further develop this definition in Section \ref{sec:narrdef}.

Building on the discussion above, we identify the following areas for improvement:
\begin{enumerate}
    \item Enhanced interpretability and contextual richness of topic labels: Labels need to provide a comprehensible, accurate description of a topic.
    \item Automation and independence of specific implementation approaches or (proprietary) tools: the label generation framework should minimize dependence on time-consuming and subjective human intervention while maintaining a high adaptability to advances in the underlying technologies.
    \item Automated validation of topic labels: Ensuring accurate and context-based labels is vital for a correct human understanding of document clusters. 
\end{enumerate}

We address the above areas by 1) conceptualizing a \emph{narrative schema} that reflects the content in each set of documents and is easy to interpret for humans. The respective narratives will then serve as topic labels. 2) We introduce NTLRAG, a retrieval-augmented generation (RAG) pipeline that operates on topic model outputs and produces reliable, human-interpretable narratives based on corpora consisting of short text documents (such as social media messages).

With our approach, we explicitly focus on short texts, as they are produced in vast amounts on different social media platforms and also present a major source for gathering individual opinions on real-world events. As user-generated short texts often only provide sparse information and a high degree of opinionated text, we employ validated news from reputable news providers to corroborate the derived narratives in a retrieval augmented generation step. Thus, we make the following contributions: 
\begin{itemize}
    \item [(i)] we establish a well-defined conceptualization of a \emph{narrative schema} for describing topic model outputs,
    \item [(ii)] we introduce a modular RAG framework tailored for structured narrative extraction and validation across multi-topic short text corpora, and
    \item [(iii)] we incorporate a dual-retriever strategy in a narrative analysis pipeline, optimized for different input source types.
\end{itemize}

To the best of our knowledge, no existing approach for topic label generation employs a retrieval-augmented generation pipeline for the creation of narrative topic labels. The remainder of this paper is structured as follows. Section \ref{sec:relwork} discusses related work. Section \ref{sec:concept} describes the NTLRAG framework and Section \ref{sec:impl} presents our NTLRAG implementation. Subsequently, Section \ref{sec:eval} discusses the evaluation of NTLRAG on three real-world datasets and a qualitative user study. In Section \ref{sec:discussion}, we discuss limitations and possible extensions, before Section \ref{sec:conclusion} concludes the paper.

\section{Related Work}\label{sec:relwork}

This section provides an overview of current practices in automatic topic labeling and evaluation (Section~\ref{subsec:autom}) as well as narrative conceptualizations in computational linguistics and information retrieval (Section~\ref{subsec:narrs}).

\subsection{Automatic Topic Labeling and Evaluation}\label{subsec:autom}

In a recent review, \cite{topiclabelingreview} classify topic labeling approaches into six main categories: (i) traditional methods (e.g.\ information retrieval), (ii) ontology-based methods, (iii) graph-based methods, (iv) human annotation, (v) hybrid approaches and (vi) approaches based on neural networks (e.g.\ transformers such as large language models). They also highlight the prevalence of domain-specific datasets for an evaluation of different methods (e.g., social media messages, reviews, etc.). Most of the approaches included in this review relied on human judgment to evaluate their results.

\cite{beyondtokenoutputs} present an approach to improve interpretability of topic labels without external sources by mapping relevant keywords to the documents in the initial corpus. Subsequently, the keywords are ranked by the number of intersecting tokens using LDA, and the top-scoring tokens are identified as topic labels. The results are evaluated through human annotation, applying three quality criteria (quality, usefulness, and efficiency) on a 5-point Likert scale. 

Another paper employs a combined approach that blends the ConceptNet \citep{Speer_Chin_Havasi_2017} knowledge graph and LLM tasks to introduce a zero-shot topic-labeling framework (Top2Label) \citep{CHAUDHARY2024122676}. The approach generates three types of topic labels (word, sentence, and summary) to provide easy-to-understand topic descriptions. 

\cite{piper-wu-2025-evaluating} use LLMs to generate topics and human-interpretable labels (keywords) from narrative texts (e.g.,\ news articles or novels) . The evaluation compares LLM-based and human-generated topic labels, indicating that the former substantially outperform the latter with respect to specific semantic characteristics. 
In a different approach, \cite{LLooM} leveraged LLMs for topic modeling and creating generalized concepts as topic labels.

To capture temporal topic evolution, a dynamic topic labeling approach is presented and automatically evaluated in \cite{GuillnPacho2024}. In \cite{nmfklabeling}, Non-Negative Matrix Factorization (NMF) has been combined with a topic model and prompt-tuning LLMs. The approach combines automatic labeling and human evaluation.

Instead of suggesting a separate topic labeling model, some studies integrate the aspect of increased topic interpretability directly into the topic modeling process. For example, the Spherical Correlated Topic Model (SCTM) \citep{SCTM} addresses the challenges of context-poor short text documents by combining word embeddings and knowledge graph embeddings. Moreover, SCTM aims to enhance interpretability of topics by accounting for the correlation between topics. However, the result is still limited to a list of representative keywords without further contextual information. 

Except for Top2Label \citep{CHAUDHARY2024122676}, all approaches discussed above provide topic labels as keyword lists, whereas NTLRAG generates context-rich narratives instead. Furthermore, most topic labeling models rely on a single information source, which is also used to create the topics themselves (e.g., \citeauthor{SCTM}, \citeyear{SCTM}; \citeauthor{GuillnPacho2024}, \citeyear{GuillnPacho2024}; \citeauthor{nmfklabeling}, \citeyear{nmfklabeling}; \citeauthor{beyondtokenoutputs}, \citeyear{beyondtokenoutputs}). In contrast, NTLRAG incorporates additional context by integrating validated news sources in a RAG step.

In evaluating topic model outputs, several studies have proposed metrics for human interpretability and human-identifiable semantic coherence. For example, \cite{NIPS2009f92586a2} conceptualize human-interpretability by developing tasks to measure human-identifiable semantic coherence. They apply the concept of word intrusion, where raters have to identify an 'intruder word' out of six terms which are presumably the most probable words for each topic. The assumption is that if the intruder word can be identified, the coherence of the topic for human readers is good. The second task is called topic intrusion. It serves as an assessment of the topic model's document composition in terms of human interpretability. Users have to identify an 'intruder topic' (represented by the eight highest probability words of this topic) among four candidate topics. 

Another conceptualization of topic interpretability  is presented in \cite{newman-etal-2010-automatic}. For evaluation, a 3-point ordinal scale is used to rate the observed coherence of a topic (3=useful (coherent), 2=neutral (partially coherent or somewhat unclear), 1=not useful (less coherent)). Here, usefulness is defined as the ability of topic-related keywords to be used in a search interface to retrieve documents about a specific topic, or the ease of finding a concise label which describes the topic.

At the model level, word intrusion and observed topic coherence provide almost identical results \citep{lau-etal-2014-machine}. \cite{evolutionoftopicmodeling} present the most prevalent metrics for evaluating topic models. Different forms of human or qualitative evaluation are discussed, such as presenting evaluators with a set of topics from multiple models and asking them to rank which one best describes each topic. 

\subsection{Narratives in computational linguistics/information retrieval}\label{subsec:narrs}

In \cite{piper-etal-2021-narrative}, the authors discuss the gap between the field of natural language processing  (NLP) and the vast amount of theoretical work on narratives within other disciplines (e.g., humanities, social sciences). They link theoretical narrative concepts to NLP applications and provide a concise definition of a narrative for practical use. Furthermore, a ten-component structure to determine the presence of narrativity is introduced: 1) teller, 2) mode of telling, 3) recipient, 4) situation, 5) agent, 6) one or more sequential actions, 7) potential object, 8) spatial location, 9) temporal specification, 10) rationale. 

As user-generated short text documents in general exhibit an irregular structure, limited context, and substantial variability compared to longer, curated texts \citep{evolutionoftopicmodeling}, we argue for an even more reduced schema for determining a narrative. \cite{Santana2023} present a review on narrative extraction from textual data. They describe it as a sub-domain of artificial intelligence (AI) that spans from retrieving information to summarizing it, extracting narrative elements from it, and producing text from this data. They suggest a narrative extraction pipeline with NLP tasks, such as Parsing, Part-of-Speech-Tagging, etc., to identify narrative components (events, participants, time, and space).

Domain-specific narrative definitions have been developed for a number of different applications. One example is the collective economic narrative \citep{collectivenarrativeeconomics}. This concept refers to a story about a topic with an economic context that is used for sense-making.

To extract narratives from user-generated short text, agent-action-target (Subject-Verb-Object) triplets have been introduced in \cite{narrativemotifs}. A survey of how different domains have conceptualized narratives for their field of research can be found in \cite{narrsurvey}. The survey also emphasizes the lack of a unanimously accepted definition of the term "narrative", suggesting that the knowledge produced by extracting narratives has a value for research that goes way beyond a mere definition of the term. 

\section{Conceptual Design}\label{sec:concept} 

In this section, we introduce the NTLRAG framework and discuss our conceptualization of narratives. NTLRAG and this conceptualization which will then be applied for deriving narrative topic labels. 

\subsection{Narrative Schema Conceptualization}\label{sec:narrdef}

In order to conceptualize narratives for creating human-interpretable topic labels from short text clusters, we build on prior definitions for narratives as 'description of a set of topic-wise interconnected documents.' We further incorporate elements from \cite{piper-etal-2021-narrative} and \cite{Herman2009} which frame narratives as experiences of human-like agents interacting with their environment, as well as from \cite{chambers-jurafsky-2008-unsupervised}, \cite{chambers-jurafsky-narrative-schemas-2009} and \cite{chambers-2013-event} which introduce narrative chains (ordered events around a protagonist) and narrative schemas (semantic role-based structures).

Based on the related work (see also Section \ref{sec:relwork}), we thus conceptualize a narrative as a data structure consisting of:
\begin{itemize}
\item[(i)] one or more actor(s); Actor $\in$ $\{$individual, group, institution, public entity, country, generic person/agent$\}$
\item[(ii)] an action associated with the actor(s); Action $\in$ $\{$verb predicate extracted from text$\}$, and
\item[(iii)] an event linking the actor(s) and the action; Event $\in$ $\{$incident, context cluster$\}$.
\end{itemize}

For improving interpretability, we further include a concise descriptive text that creates a sentence from the three narrative elements mentioned above. Subsequently, we use this four-element structure as a narrative schema for topic labels extracted and generated by NTLRAG.

\subsection{NTLRAG}

\begin{figure*}[htbp]
\includegraphics[scale=0.147]{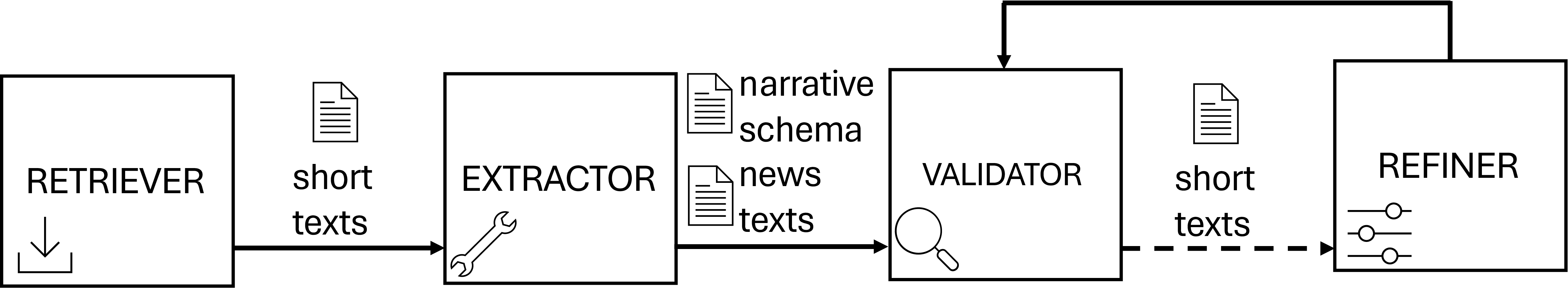}
\caption{NTLRAG pipeline: Retriever, Extractor, Validator, and the conditional Refiner.}
\label{fig:NTLRAGpipeline}
\end{figure*}

NTLRAG is a retrieval-augmented generation (RAG)-based framework that is designed for working on the output produced by a topic model. The framework is inherently flexible, allowing for the integration of different RAG implementations and components. Additionally, through its modular structure, extensions or adaptations can easily be implemented. Figure \ref{fig:NTLRAGpipeline} presents a high-level view of the NTLRAG pipeline, a detailed pseudo-code specification is found in Algorithm~\ref{ntlragpipeline}.

The input to the RAG pipeline is the query which is provided by the keyword list that has been created by the underlying topic model. Short text documents are represented as $D_S$, while $D_N$ denotes curated and trustworthy news content. Extracted narratives ($NarrativeSchema$) are processed through the validation function and re-extracted in the refine function if they are assigned to the 'refine' category (for details see below).
\bigskip

\begin{algorithm}[H]
\caption{NTLRAG - Pipeline}
\label{ntlragpipeline}
\KwIn{Keyword-based query $q$}
\KwOut{Final Output $NarrativeSchema$}

$D_S \gets \text{RETRIEVERS}(q)$\;
$D_N \gets \text{RETRIEVERN}(q)$\;
$D \gets \text{Concatenate}(D_S, D_N)$\;

$M \gets \text{EXTRACTOR}(\text{prompt=EXTRACT}, \text{input}=D_S)$\;

\Repeat{$c = \text{"approve"}$}{
    $c \gets \text{VALIDATOR}(\text{prompt=VALIDATE}, \text{inputs}=[NarrativeSchema, D])$\;

    \If{$c = \text{"refine"}$}{
        $NarrativeSchema \gets \text{REFINER}(\text{prompt=REFINE}, \text{inputs}=[NarrativeSchema, D_S])$\;
    } 
}

\Return $NarrativeSchema$\;
\end{algorithm}

\bigskip
\textit{RETRIEVER.} The RAG pipeline includes four main steps implemented via four software components (see also Algorithm \ref{fig:NTLRAGpipeline}). The first component in this pipeline is the Retriever which focuses on document retrieval. In this step, relevant documents are collected from two sources: the topic model output and a corpus of related news documents. The topic model output must include the following information: 
\begin{itemize}
    \item[(i)] the full text of each document,
    \item[(ii)] its assigned topic number, and
    \item[(iii)] a set of keywords per topic that are essential for building the query for retrieval.
\end{itemize}

The news document corpus comprises texts that closely align with the content of the short-text documents. For example, when analyzing social media messages related to a specific incident, additional documents  include trustworthy news reports on the same event (e.g., a shooting, a flood, or a wildfire). 

The framework remains source-independent and can be used with short text corpora from any social media platform as well as news articles from any reputable news provider. Regarding news articles, NTLRAG requires only the article text, title, or a concise summary of its content. In general, full articles are preferred for news context documents. Depending on the specific implementation of the Extractor component, complete news articles could be too large, though. For example, different large language models (LLMs) may impose different limitations on the length of accepted input sequences. Therefore, using summaries or even just titles can serve as alternatives. In any case, for both corpora (news and short text), optional metadata (e.g., publication date, news outlet) can be incorporated to facilitate downstream analysis and interpretation of the results.

\begin{figure}[ht]
\includegraphics[scale=0.25]{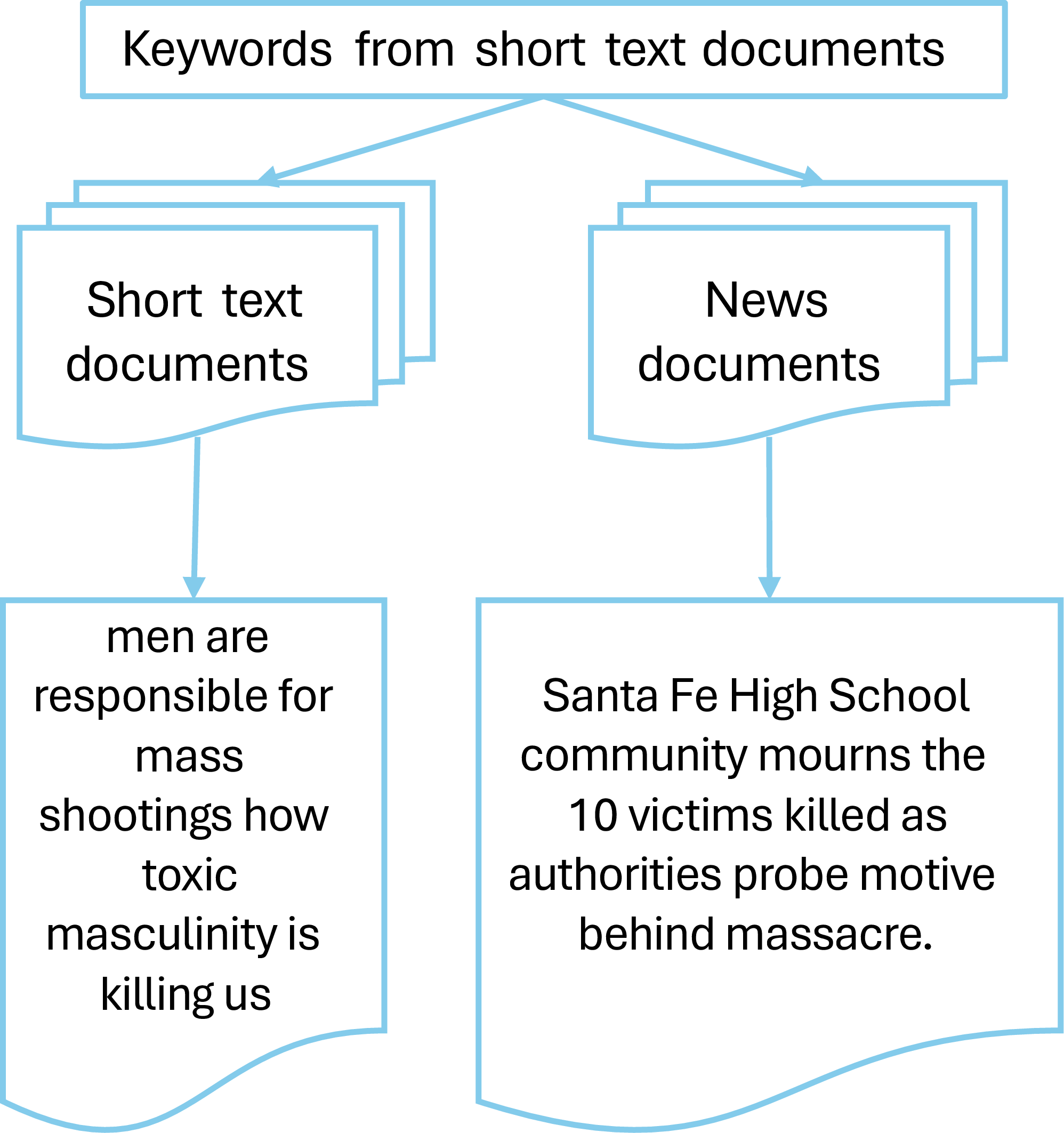}
\caption{RETRIEVE step including an example of retrieved context documents.}
\label{fig:retrieve}
\end{figure}

Distinct retrieval functions are used for each corpus; however, the query remains identical for both contexts and is derived from the keywords generated by the topic model.

\textit{EXTRACTOR.} The second step involves the extraction of a narrative description from the retrieved short text documents. Because NTLRAG relies on a RAG architecture, one popular choice for the Extractor component is using an LLM. However, in general the Extractor can be implemented using any suitable component or technology.

The output of the Extractor is based on a structured data model with five elements: 
\begin{itemize}
    \item [(i)] a topic identifier, 
    \item [(ii)] the actor(s) involved,
    \item [(iii)] an action,
    \item [(iv)] an event linking the actor(s) and the action, as well as
    \item [(v)] a concise one-sentence narrative summary referred to as \emph{narrative topic description}.
\end{itemize} 

This structured representation ensures consistency, comparability, and facilitates subsequent validation steps. When the Extractor is implemented based on an LLM, prompt engineering is crucial to prevent hallucinations or irrelevant responses. The proposed prompt instructions that we use in our implementation are detailed in Section \ref{sec:impl}. 

\begin{figure}[htbp]
\includegraphics[scale=0.25]{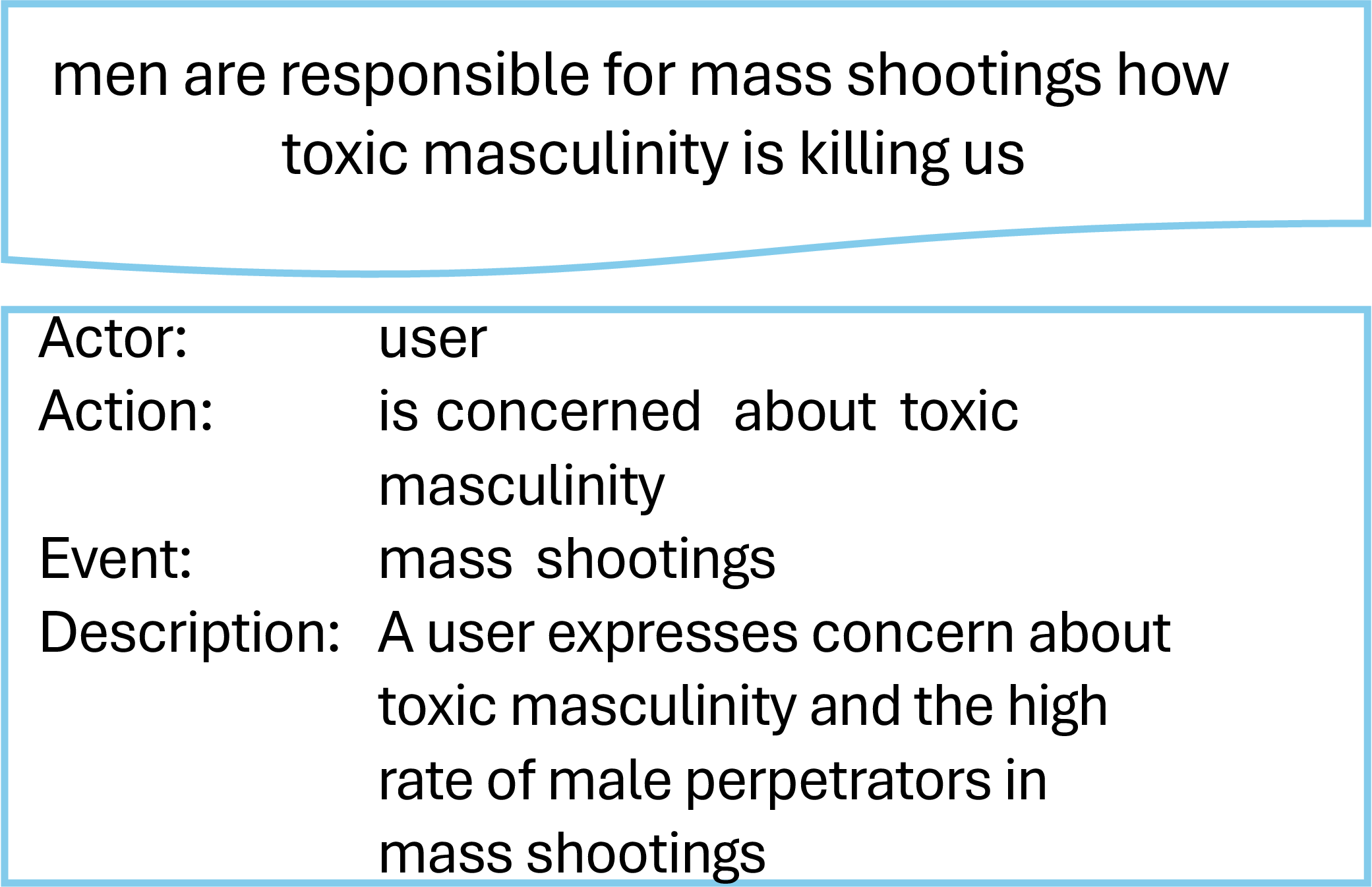}
\caption{EXTRACT step with an example short text document and the derived narrative data structure.}
\label{fig:extract}
\end{figure}

\textit{VALIDATOR.} Once a narrative topic description has been derived from the corresponding short text documents, it subsequently undergoes a validation process. 

\begin{figure}[ht]
\includegraphics[scale=0.25]{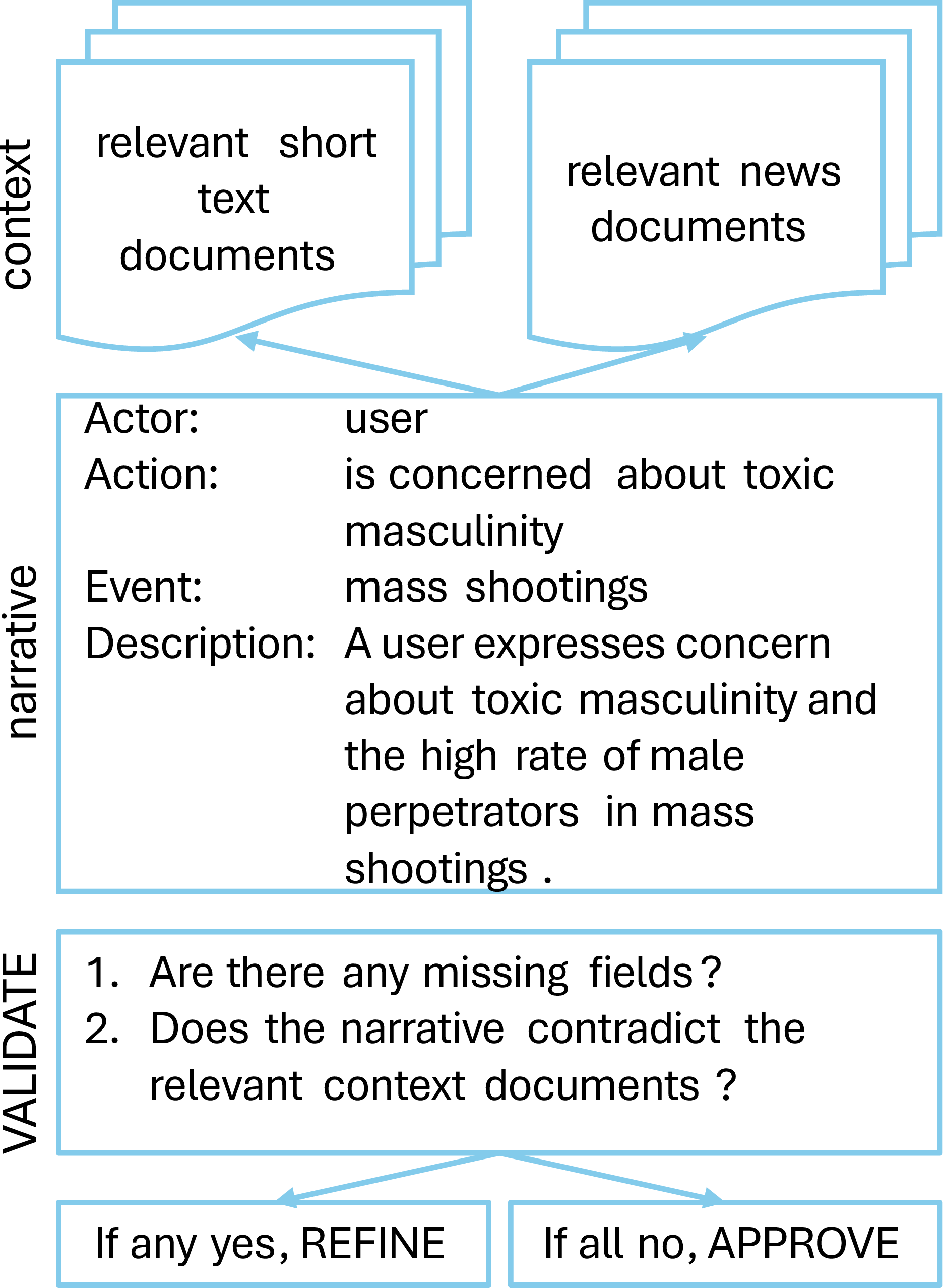}
\caption{VALIDATE step with an example narrative and main categorization criteria.}
\label{fig:validate}
\end{figure}

For this step, the short text and news documents serve as contextual references. As shown in Figure \ref{fig:validate}, two main criteria determine the validation outcome. First, a validity check verifies whether each attribute of the \emph{narrative schema} (actor, action, event, and description) exists. Second, the Validator assesses the narrative's quality. It is straightforward, to implement this component based on an LLM, however any suitable technology can be used for the Validator. Subsequently, assessment will be performed through retrieval augmented generation, using a tailored prompt and relevant documents. The primary quality criterion is consistency with the context documents, enforced through a “non-contradiction” clause. For our implementation, iterative prompt engineering indicated that a negative detection approach is most effective (for details see Section \ref{sec:impl}). Therefore, all narratives are initially assigned an “approved” category which is revised only if
\begin{itemize}
\item[(i)] one or more elements (actor, action, event, description) are missing, or
\item[(ii)] the narrative contradicts the context documents.
\end{itemize}

In this paper, a \emph{contradiction} refers to an event or state that occurs in both the narrative and the context documents, but is described in conflicting ways. For example, the narrative “actor: farmers; action: protest; event: new regulations on cattle care; description: farmers protest new regulations on cattle care in Wyoming” contradicts the context “Farmers supported the adoption of new regulations specifying cattle care requirements in the U.S.”. In contrast, differences in granularity, such as specifying a more precise location (Wyoming instead of the United States), are not considered contradictions. 

In addition, the prompt includes further guidance (e.g., 'Do not consider tone or language when grading the narrative') and explicit constraints to prevent hallucinations. If a narrative is assigned the 'refine' category, it is moved to the Refiner component, where a new narrative is extracted. Each narrative is re-assessed until it is assigned to the 'approved' category. At this point, a chain-of-thought element is part of the pipeline, forcing the Validator to explain the assigned category (see examples in Section~\ref{sec:impl}). 

\textit{REFINER.} In this step, a new narrative is extracted from the short text documents. The extraction procedure is identical to that used by the Extractor, but implemented as a separate function to allow flexibility in sourcing narratives from both short texts and news documents.

\section{NTLRAG Implementation}\label{sec:impl} 

\begin{figure}[htbp]
\includegraphics[scale=0.155]{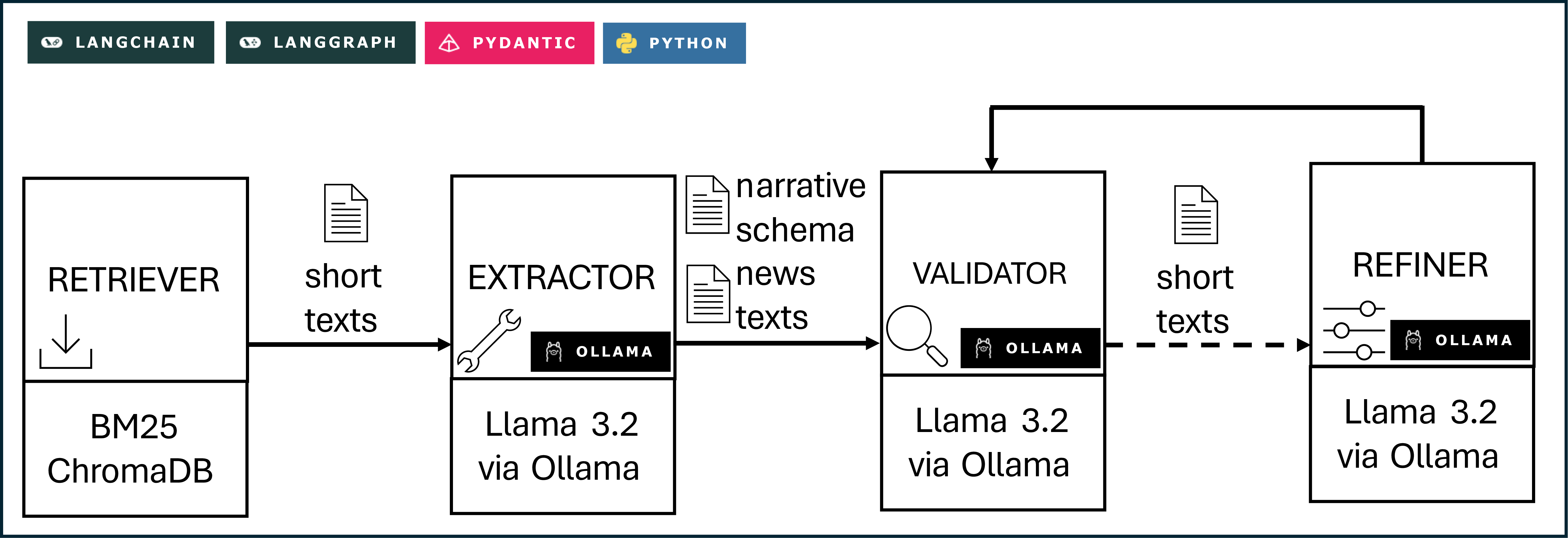}
\caption{NTLRAG components with implementation choices. LangChain, LangGraph and Pydantic for general RAG orchestration and Llama LLM via Ollama for Extractor, Validator and Refiner.}
\label{fig:NTLRAGpipelineComponents}
\end{figure}

We provide a full implementation of the conceptual framework presented in Section \ref{sec:concept}. While we evaluated multiple different options and configurations, this section describes one particular implementation of NTLRAG. Figure~\ref{fig:NTLRAGpipelineComponents} presents the implementation choice for each component. Our implementation is publicly available for download.\footnote{See \textit{\url{https://github.com/lisagrobels/NTLRAG}.}}

For our implementation, we used LangChain and the LangGraph framework \citep{Chase_LangChain_2022} implemented in Python. NTLRAG can be described as an orchestrated RAG pipeline with a loop that is dependent on the outcome of the Validator (see also Algorithm \ref{ntlragpipeline}). In the development process, we ensured that the model performs efficiently without requiring specialized computational infrastructure.\footnote{Runtime environments for testing our implementation were chosen based on each task’s requirements: for data preprocessing, we used the standard CPU runtime (RAM: 51 GB), while topic modeling and NTLRAG execution employed T4 GPUs (GPU RAM: 15 GB, CPU RAM: 51 GB) and A100 GPUs (GPU RAM: 40 GB, CPU RAM: 83.48 GB) interchangeably, depending on availability. The LLM implementation was set up using Ollama and the Llama 3.2 model (3B parameters) \url{https://ollama.com/library/llama3.2}.}

\textit{RETRIEVER.} As outlined in Section~\ref{sec:concept}, the first step of NTLRAG consists of retrieving relevant documents based on topic model output (short text documents) as well as a news corpus. The keywords produced by the upstream topic model are used as the query to retrieve information from both sources. The framework supports a wide range of retriever and data storage configurations. Lexical approaches (e.g., term frequency-based such as BM25) and vector-based approaches can be used independently or integrated within an ensemble retriever \citep{afzal2024optimizingretrievalaugmentedgeneration}. 

In our implementation, we applied BM25 (Best Matching 25) for the short text corpus. BM25 is a bag-of-words model that ranks documents based on TF-IDF scores and document length \citep{robertson1995okapi}. This choice is motivated by the fact that the query strings are generated from the short text corpus itself which reduces the likelihood of mismatches arising from lexical variation. Our BM25 retriever for the short text corpus is organized as a dictionary. It maps each topic to a dedicated BM25 instance, enabling NTLRAG to iterate over topics efficiently.
 
For news documents, we used a vector-based retrieval method, whereby the documents are first transformed into word embeddings before the retrieval function is applied. The matching and ranking process is performed based on cosine similarity scores. We store and index embeddings using ChromaDB \citep{chroma2025}. Retrieval queries are static and comprise relevant keywords for each topic as determined through topic modeling (see Tables~\ref{tab:example_validate1} and \ref{tab:example_validate2} for example queries).

\textit{EXTRACTOR.} In our implementation, the extractor component is based on an LLM. To align with the \emph{narrative schema} defined in Section \ref{sec:narrdef}, employing an Extractor that supports structured output is recommended (such as a suitable LLM). We use Llama 3.2 via Ollama and Pydantic \citep{Colvin_Pydantic_Validation_2025} to validate and guide the LLM's answers. The corresponding data structure is shown in Table~\ref{tab:pydantic}. Furthermore, we leverage the structured output method provided by LangChain and Ollama.\footnote{\url{https://python.langchain.com/docs/how_to/structured_output/\#typeddict-or-json-schema}, \url{https://ollama.com/blog/structured-outputs}.} 

\begin{table*}[h]
\centering
\caption{NTLRAG Pydantic model schema for narratives.}
\begin{tabular}{llp{9cm}}
\toprule
\textbf{Field} & \textbf{Type} & \textbf{Description} \\
\midrule
topic\_id & String & The topic ID of the narrative \\
actor & String & The actor(s) of the narrative \\
action & String & Action that is carried out by actor(s) or other entities or individuals\\
event & String & The event linking the actor(s) and their action \\
description & String & A one sentence long description of the narrative \\
\bottomrule
\end{tabular}
\label{tab:pydantic}
\end{table*}   

The Pydantic JSON schema description extends the prompt used for narrative extraction and enables efficient validation of output and easy use in downstream tasks. 

\begin{table*}[h!]
\centering
\caption{Example in- and output of the EXTRACT and VALIDATION step in the \textit{approved} category.}
\begin{tabular}{|l|p{9cm}|} 
\hline
& \textbf{Output} \\ \hline
Query               & largest norm shootings modern biggest deadliest lifetime frequency worst proposals \\ \hline
Retrieved documents & we’ve seen the biggest mass shooting in history headline too many times in my lifetime. 
                      710 deadliest mass shootings happened in my lifetime; my home state holds the most. 
                      This is the deadliest mass shooting in US history — where’s the outrage, where’s the policy proposals? 
                      Mass shootings are an American norm, sad but true. \\ \hline
Narrative           & \textbf{actor:} user, \textbf{action:} expresses frustration with gun violence, \textbf{event:} mass shootings. 
                      \textbf{description:} The user expresses frustration with mass shootings in the US, highlighting their increasing frequency and casualty count. \\ \hline
Validation          & The narrative is consistent with the context and does not contradict any information provided. \\ \hline                        
\end{tabular}
\label{tab:example_validate1}
\end{table*}

The composition of the prompt strongly influences the quality of the LLM output. The final prompt was developed through an iterative process involving the addition, removal, and refinement of prompt components. It comprises three main elements: (i) the system message, e.g., ‘You are an information extraction system...'; (ii) the extraction rules, e.g., ‘STRICTLY use only information found in the provided documents…’; and (iii) the retrieved short text documents. Within the extraction rules, the Pydantic data model was further specified. For instance, the ‘user’ value was specified as valid for the ‘actor’ field when the LLM could not identify any other actor(s). During validation, this value is always accepted provided that all other fields are valid.

Also, the LLM was prompted to disregard any differences in tone and language. The complete prompt is provided in \ref{app1}, and an example of retrieved documents, including the corresponding query, narrative output, and validation explanation, is shown in Tables~\ref{tab:example_validate1} and~\ref{tab:example_validate2}. After successful extraction the narrative is forwarded to the Validator.  

\textit{VALIDATOR.} For narrative validation, the LLM is provided with the output from the extraction step and instructed to categorize each narrative as either ‘approve’ or ‘refine.’ Narratives categorized as ‘approve’ are considered final and do not undergo further processing, whereas those categorized as ‘refine’ are forwarded to the refinement step. In particular, the validation step draws on two context corpora (short text documents and news articles) along with a prompt template. A detailed description of the prompt logic is provided in Section~\ref{sec:concept}, and the full prompt is included in \ref{app1}. An integrated chain-of-thought component generates an explanation for each LLM validation decision. Examples of queries, retrieved documents (retrieval), extracted narratives (extract), validation outcomes (validate), and refined narratives (refine) are presented in Tables~\ref{tab:example_validate1} and \ref{tab:example_validate2}.

\begin{table*}[h!]
\centering
\caption{Example in- and output of the EXTRACT and VALIDATION step in the \textit{refine} category.}
\begin{tabular}{|l|p{9cm}|} 
\hline
& \textbf{Output} \\ \hline
Query               & ar15 ar15s foundno shotgun ar rifle pistol used ar15style revolver  \\ \hline
Retrieved documents & no ar15 foundno ar15 foundno ar15 foundno ar15 foundno ar15 foundno ar15 foundno ar15 foundno ar15 foundno
                     he was armed with an ar15style rifle a pistol a shotgun standard bill of fare these days.
                     hey the attacker was armed with an ar15style rifle a pistol a shotgun and pipe bombs maybe you should...
                     no ar15 at santa fe shooter used a pistol and shotgun.
                     he used a shotgun and pistol not a ar15 still tragic. \\ \hline
Narrative I          & \textbf{actor:} user, \textbf{action:} attacked, \textbf{event:} office building. 
                      \textbf{description:} Attacker armed with an AR-15 style rifle and other weapons carried out the attack. \\ \hline
Validation (refine)  & The narrative includes hallucinations (i.e. facts not present in the context). The context does not mention  'office building' or the action 'attacked'. Also, no ar15 found. \\  \hline
Narrative II         & \textbf{actor:} user, \textbf{action:} attacked, \textbf{event:} Santa Fe High School 
                      \textbf{description:} The attacker used a shotgun and pistol to attack Santa Fe High School. \\   \hline
Validation (approve) & The narrative is consistent with the context, as it reports that the attacker used a shotgun and pistol to attack Santa Fe High School. The actor 'user' is also valid in this scenario. \\ \hline         
\end{tabular}
\label{tab:example_validate2}
\end{table*}

Each narrative is allowed a maximum of 100 ‘refine’ iterations, as iterative testing indicated no further improvements beyond this point. This limit can be adjusted depending on the specific use case, of course. However, increasing the retry limit requires a corresponding increase in the recursion limit for LLM calls, which in turn results in a greater number of LLM requests and potentially higher cost for online models with request-based pricing.

\textit{REFINER.} This step mirrors the functionality of the Extractor, providing short text context documents to the LLM to populate the \emph{narrative schema}. The prompt from the Extractor remains unchanged. Since the same context documents are used, the resulting narrative content is not expected to deviate substantially from the extraction.

\section{Evaluation}\label{sec:eval}

In this section, we report on the results obtained from applying our approach to three real-world datasets consisting of more than 6.7 million social media messages that have been sent by more than 2.7 million users. Furthermore, the resulting narratives are evaluated through a questionnaire-based human assessment.

\subsection{Datasets and Preprocessing}

In line with numerous studies in this field \citep{topiclabelingreview}, we employed domain-specific datasets to evaluate our framework. The datasets are summarized in Table \ref{tab:datasets} and have partially been used in other topic modeling approaches in \cite{snams2022}, \cite{complexis22}, \cite{emotion-exchange-motifs-ieee-ic} and \cite{emotion-exchange-motifs-chb}. 

\begin{table*}[ht]
\footnotesize
\caption{Three datasets used in the case study.}
\begin{tabular}{llllc} 
\toprule
\textbf{Location} & \textbf{Observation period} & \textbf{Users} & \textbf{Tweets} & \textbf{Unique Tweets}  \\ 
\midrule
Las Vegas (NV)         & 2-14 October 2017     & 1,394,070      & 3,436,187       &  505,850 \\ 
Santa Fe (TX)           & 18-25 May 2018    & 458,644        & 967,674         & 113,146     \\ 
El Paso (TX)            & 3-18 August 2019    & 939,940        & 2,307,577       &  318,368   \\
\bottomrule
\end{tabular}
\label{tab:datasets}
\end{table*}

\begin{itemize}
    \item The Las Vegas dataset refers to a mass shooting at the Route 91 Harvest music festival on October 1, 2017. A shooter killed 60 people and wounded at least 413 others. The subsequent panic increased the total number to 867 injuries \citep{lasvegasinfo}.
    \item The Santa Fe dataset refers to a school shooting event that happened in May 2018 in Santa Fe, Texas. Eight students and two teachers were shot dead and ten others wounded \citep{safeinfo}.
    \item The El Paso dataset refers a shooting in a Walmart in El Paso, Texas on August 3, 2019. In this incident, 23 people were killed, another 22 have been wounded \citep{elpasoinfo}.
\end{itemize}

The datasets were collected via Twitter's API with academic access during the observation periods referred to in Table \ref{tab:datasets}. Only unique tweets were retained for topic modeling. Pre-processing steps included the removal of retweet identifiers, user mentions, special characters, URLs, and hashtags. Following the procedure outlined in \cite{grootendorst2022bertopicneuraltopicmodeling}, all text messages have been converted to lowercase.

To obtain document clusters, we applied BERTopic \citep{grootendorst2022bertopicneuraltopicmodeling}, a modular framework based on document word embeddings. The dimensionality of the word vectors is reduced using Uniform Manifold Approximation and Projection (UMAP). Embeddings are then grouped using a density-based clustering algorithm (HDBSCAN, \citeauthor{hdbscan}, \citeyear{hdbscan}). For our datasets, the model was configured through hyperparameter tuning via the  Optuna framework \citep{optuna_2019}. We computed a weighted score based on the Silhouette Score, Diversity, as well as the Outlier Ratio to determine the optimal number of minimum clusters for HDBSCAN and the minimum number of documents per topic for BERTopic itself. This procedure resulted in 2,335 topics for the Las Vegas dataset, 948 topics for the El Paso dataset, and 294 for the Santa Fe dataset.

\subsection{NTLRAG Evaluation and User Study}

A manual inspection of the model's results has already revealed insights into NTLRAG's superior ability to handle negated statements in comparison to keywords. Table~\ref{tab:example_validate2} shows a corrected narrative that initially included the information that a perpetrator was using an AR15 rifle. However, the related context documents did not support this information; therefore, NTLRAG refined the narrative and added facts. In contrast, the keywords alone ('ar15 ar15s foundno shotgun ar rifle pistol used ar15style revolver') do not give any meaningful information about this incident. 

Another example demonstrating NTLRAG's superiority to keywords is the added context it provides. In Table~\ref{tab:example_context}, we present retrieved short text documents, their corresponding narratives, and associated keywords. Whereas the keywords do not provide any connection to the shooter, the corresponding narrative adds him to the context, making the information considerably easier to interpret.

\begin{table*}[h!]
\centering
\caption{Example in- and output of the EXTRACT step}
\begin{tabular}{|l|p{9cm}|} 
\hline
& \textbf{Output} \\ \hline
Retrieved documents & las vegas gunman described as welloff gambler peaceful.
                     las vegas gunman described as welloff gambler and a loner.
                     killers gambling habits revealed.
                     mysteriously calculated gambling habits.
                      las vegas shooter gambled 100000 an hour at video poker. \\ \hline
Narrative (NTLRAG) & \textbf{actor:} user, \textbf{action:} described, \textbf{event:} gambler. 
                      \textbf{description:} A well-off gambler, described as peaceful and a loner, carried out the attack. \\ \hline
Keywords (BERTopic) & gambling gambler poker compulsive gambled transactions habits gamble stakes welloff \\  \hline
\end{tabular}
\label{tab:example_context}
\end{table*}

For a human evaluation of NTLRAG-generated narratives in comparison to traditional lists of keywords, we implemented an application via Shiny \citep{shiny} that was hosted on shinyapps.io.\footnote{\url{https://www.shinyapps.io/}.} We designed an evaluation form comprising an introduction, an example page with three examples how to perform the rating procedure, and an evaluation page. Screenshots of the evaluation application are provided in \ref{app2}.

Because NTLRAG uses standard topic model output to generate narrative topic descriptions, our goal is to evaluate the human interpretability of narrative topic descriptions, rather than the quality of the underlying topic model output. To this end, we developed a procedure to determine how well the narrative topic description can summarize documents in each cluster. The evaluation included 50 topics, where each topic is described by (i) the narrative, consisting of the actor, action, event triplet, and a description, as well as (ii) the keywords as produced by BERTopic via class-based-TF-IDF weighting.  

NTLRAG was evaluated for its usefulness (see, e.g., \citeauthor{lau-etal-2014-machine}, \citeyear{lau-etal-2014-machine}; \citeauthor{beyondtokenoutputs}, \citeyear{beyondtokenoutputs}; \citeauthor{lau-baldwin-2016-sensitivity}, \citeyear{lau-baldwin-2016-sensitivity}) on a 3-point-ordinal scale to rate topic labels as suggested by \cite{lau-etal-2014-machine}, \cite{lau-baldwin-2016-sensitivity} and \cite{mimno-etal-2011-optimizing}. Raters were provided with a set of documents belonging to the same cluster, a narrative label describing the cluster (including actor, action, event, and description), and for comparison a list of keywords automatically generated by the BERTopic model (see also \ref{app2}).

In our evaluation, usefulness consisted of two main aspects:

\begin{itemize}
    \item \textit{Accuracy} – Does the label reflect the content of the documents?
    \item \textit{Clarity} – Can you easily understand the cluster's overall message from the label? 
\end{itemize}

Raters had to evaluate 50 narratives and keyword lists which were randomly selected from all three datasets. Overall, 16 human raters participated in the survey with an average time effort of 50 minutes each. Participants were all familiar with social media texts, two were considered experts as they work with user-generated short texts on a regular basis.  

\begin{table*}[h!]
\centering
\caption{NTLRAG narratives and BERTopic keywords examples. Three NTLRAG narratives with the highest human rating and the corresponding BERTopic keyword lists.}
\begin{tabular}{|p{1.6cm}|p{2cm}|p{2cm}|p{3.5cm}|p{2.5cm}|}
\hline
\multicolumn{4}{|c|}{\textbf{Narratives}} & \textbf{Keywords} \\ \hline
ACTOR & ACTION & EVENT & DESCRIPTION &  \\ \hline
Jason Aldean & resume touring & Las Vegas shooting, & Country star Jason Aldean resumed his tour in Tulsa after the Las Vegas shooting marred his performance. & resumes tour resume resuming marred resumed tulsa touring tuls star \\ \hline
Natca Air Traffic Controller & warned & Airport During Las Vegas Shooting & An air traffic controller at a concert in Las Vegas warned airport authorities about the shooting. & maher controller rips traffic snarkily natca accuratebill airport shootg warned \\ \hline
NRA & opposes & US ban on gun devices & The NRA opposes an outright US ban on gun devices used by Las Vegas killer. & outright opposes devices killerend ampvisitors zou sci accomplished newtop killer\\ \hline
\end{tabular}
\label{tab:highestnarr}
\end{table*}

\begin{table*}[h!]
\centering
\caption{NTLRAG Narratives and BERTopic keywords examples. Three BERTopic keyword lists with the highest human ranking and the corresponding NTLRAG narratives.}
\begin{tabular}{|p{1.6cm}|p{2cm}|p{2cm}|p{3.5cm}|p{2.5cm}|}
\hline
\multicolumn{4}{|c|}{\textbf{Narratives}} & \textbf{Keywords} \\ \hline
ACTOR & ACTION & EVENT & DESCRIPTION &  \\ \hline
user & criticizing & the user & The user is criticizing a person for lying. & idiot stupid liar lying moron ignorant dumb lies stupidity fool \\ \hline
Dallas Cowboys & donating &  El Paso Victims Relief Fund & The Dallas Cowboys NFL Foundation donated \$50,000 to support those affected by the El Paso shooting. & foundation charity fund norte legends relief donated proceeds 50000 donating \\ \hline
[name of perpetrator] &  posted and shared & Neonazi imagery & [name of perpetrator] posted neonazi imagery online before killing at least eight people. & imagery neonazi online posted onlinebefore onlin himgtdimitrios suspct imgry onlineshared \\ \hline
\end{tabular}
\label{tab:highestkeyw}
\end{table*}

To assess inter-rater reliability, we first calculated Krippendorff's Alpha (see, e.g., \citeauthor{beyondtokenoutputs}, \citeyear{beyondtokenoutputs}; \citeauthor{khaliq-etal-2024-ragar}, \citeyear{khaliq-etal-2024-ragar}; \citeauthor{Santana2023}, \citeyear{Santana2023}). 
The alpha coefficient \citep{krippendalpha} ranges from -1 (systematic disagreement) to 1 (perfect agreement), with 0 indicating no agreement. 

For the narrative ratings in our evaluation, the alpha values ranged from 0.376 (Santa Fe dataset) to 0.397 (Las Vegas dataset). Participants did not require prior experience in topic modeling or short text data analysis. However, we found that expert participants reached an inter-rater agreement of 0.523 across all datasets, while non-experts reached an agreement of 0.393. While alpha values for narratives can be interpreted as 'Fair Agreement' ($0.2 <= \alpha <= 0.4$) \citep{hughes2021krippendorffsalpharpackagemeasuring}, the main focus of our interpretation is the comparison between narratives and keywords for topic labeling.    

\begin{figure*}[htbp]{\centering
\includegraphics[scale=0.7]{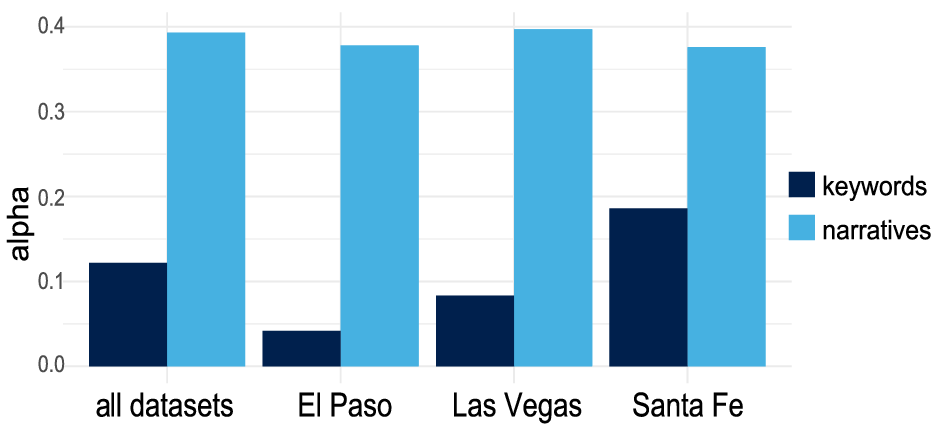}
\caption{Krippendorff's alpha for narratives and keywords of each dataset and all three datasets combined.}
\label{fig:interrater}}
\end{figure*}

Figure~\ref{fig:interrater} shows Krippendorff's alpha values for narratives and keywords separately for each dataset as well as the combined dataset. The notable difference between narrative and keyword scores suggests a significant gap in human ability to understand the central message of a document cluster. We follow \cite{beyondtokenoutputs} in their interpretation of a large gap between alpha values of two categories and argue that, based on these results, evaluators found interpreting keywords more difficult than the corresponding narratives. Table~\ref{tab:highestnarr} showcases the top three narratives with the highest average ratings, and Table~\ref{tab:highestkeyw} presents the keyword lists with the highest ratings and their corresponding narratives.

Overall, NTLRAG-generated narratives received an average rating of 2.467, whereas ordinary BERTopic keyword lists received an average rating of 1.61 (both on the 1-3 scale). Out of the evaluated 50 narratives, 49 received at least once the highest ('useful') rating. For the keyword lists, the corresponding value was 33. In 94.73\% of all cases, the narrative received the same or a higher rating than the keyword list. For 63.25\% of all cases, evaluators strictly preferred the narrative description.  

\section{Discussion}\label{sec:discussion} \subsection{Conceptual Design}

NTLRAG produces context-rich topic labels (narratives) that outperform traditional keyword lists in terms of human interpretability (see Section \ref{sec:eval}). Compared to existing approaches, it offers a more comprehensive description of topics without requiring user intervention (see also Section \ref{sec:impl}). 
NTLRAG refines a narrative until it is accepted or a refinement limit is reached. Depending on the specific document cluster or on how the refinement limit is chosen, this might result in low-quality narratives. Further improvements could be achieved through a human-in-the-loop approach, such as integrating an evaluation tool into the pipeline and converting validation from fully-automatic to semi-automatic. Additionally, if NTLRAG is combined with a multinomial topic model, the retrieval step could incorporate document weighting to reduce the likelihood of retrieving low-probability context documents.

\subsection{Implementation}

Our NTLRAG example implementation also comes with certain limitations. All general drawbacks associated with using LLMs for generative tasks also apply to our NTLRAG implementation, particularly the high dependence on an LLM's training data, vulnerability to hallucinations, and limited reproducibility. This may affect the validity and interpretability of results. However, the NTLRAG pipeline and implementation (see Sections \ref{sec:concept} and \ref{sec:impl}) are designed to address many of these challenges. In particular, we use a RAG-based design, chain-of-thought components, and validation steps. Nevertheless, the incorporated LLMs may still produce partially incorrect or irrelevant outputs. Since the choice of a particular LLM significantly influences both the results and the required prompt structure, we recommend defining selection criteria (e.g., latency, cost) before formulating the full prompt. Remarks from our human evaluators suggested that particular document clusters had too specific descriptions, which, again, might result in underrepresented or missing topical aspects of the cluster. 

\section{Conclusion}\label{sec:conclusion} 

In this paper, we introduced a novel topic labeling framework that can operate on the output of any classical topic model. NTLRAG is modular and can be implemented using a wide range of methods for text extraction (e.g., LLMs) and orchestration frameworks, requiring minimal human effort. 
Moreover, we introduce a 'narrative schema' that serves as a context-rich option for labeling and describing topics. In our evaluation with human users, we found that the narrative schema shows superior usability for interpreting document clusters (i.e., topics) compared to traditional keyword lists. 

In particular, we conducted a user study where 16 human evaluators found that NTLRAG narratives are straightforward to comprehend and that narrative topic labels more effectively represented the underlying document clusters as compared to traditional keyword lists. An implementation of NTLRAG is publicly available for download.\footnote{\url{https://github.com/lisagrobels/NTLRAG}.}  Among other things, we tested our example implementation in Google Colab\footnote{Google Colab provides free access to computing resources \url{https://colab.google/}}, which already provides sufficient computational resources for our implementation. In future research, we plan to incorporate a human-in-the-loop component to further mitigate the risk of LLM hallucinations and increase validity. Additionally, further tests to compare results with different news sources and lengths (titles vs.\ full-text) are planned in order to determine the minimal amount of validated news data that is required for our approach. 

\section*{Declaration of competing interest}
The authors declare that they have no known competing financial interests or personal relationships that could have appeared to influence the work reported in this paper.

\section*{CRediT authorship contribution statement}
\textbf{Lisa Grobelscheg:} Conceptualization, Methodology, Software, Validation, Formal analysis, Investigation, Resources, Writing - original draft, Writing - review \& Editing Visualization, Software, Methodology, Data curation, Validation. \textbf{Ema Kahr:} Conceptualization, Methodology, Software, Validation, Investigation, Supervision, Data Curation, Writing - original draft, writing - review \& editing. \textbf{Mark Strembeck:} Conceptualization, Methodology, Writing - Review \& Editing, Visualization, Supervision.

\section*{Funding Statement}
This research received no external funding.

\section*{Ethical Approval Statement}
The study adhered to the ethical principles of the Declaration of Helsinki. All participant information was handled confidentially, and data privacy was strictly maintained throughout the study.

\section*{Declaration of AI using Assisted Technologies}
During the preparation of this work, the authors used Grammarly and ChatGPT to improve English clarity and conduct spelling and grammar checks. After using these tools, the authors reviewed and edited the content as needed and take full responsibility for the content of the published article.

\section*{Data Availability Statement}
Regarding the datasets used in our case studies, we will share only insights derived from them due to restrictions imposed by the social media platform (X, formerly known as Twitter).

\section*{Code availability}
The code for our implementation is publicly available on GitHub.\footnote{\url{https://github.com/lisagrobels/NTLRAG}}

\appendix
\section{Prompts} \label{app1}

Prompt for narrative extraction: \\
You are an information extraction system.\\
Your task:\\
From the following documents, extract ONLY the information present to fill the following JSON object:\\
{{\\
   'actor': '',\\
   'action': '',\\
   'event': '',\\
   'description': ''\\
        }}\\
Rules:\\
    - STRICTLY use only the information found in the provided documents.\\
    - Absolutely NO external knowledge, assumptions, or inferred details.\\
    - Your output will be discarded if it contains information not directly from the documents.  \\  
    - 'action' should include at least one verb.\\
    - 'event' is the object of the action and can include nouns and noun phrases.\\
    - 'actor' can be any entity or multiple entities (individual, group, institution, public entity, country, etc.).\\
    - ONLY if you cannot determine an 'actor', use 'user'.\\
    - 'description' must summarize the narrative in one sentence and must be consistent with 'actor','action' and 'event'.\\
    - Output ONLY the JSON object, nothing else.\\

      DOCUMENTS:\\
       -------------------
       
        '''  )

Prompt for narrative validation:\\
You are a narrative fact-checker. Your task is to analyze a narrative in the context of supporting documents and determine if it is consistent.\\

    Rules for Labeling\\

    Start by assuming the narrative is **approved**. Change it to **refine** only if:\\

    1. The narrative **contradicts** the context (i.e. directly conflicts).\\
    2. The narrative includes hallucinations (i.e. facts not present in the context).\\

    Approve if:\\
    - The narrative is CONSISTENT with the context.\\
    - The narrative does not contradict the context (i.e. tells the opposite).\\
    - Approximate matches exist (e.g. 'America' and 'US').\\
    - The actor is 'user' (this is always valid and must be **approved** if other fields are valid).\\

    Do NOT:\\
    - Guess or invent information.\\
    - Consider grammar, tone, or style.\\
    - Penalize narratives that are vague but not contradictory.'''\\

    'Use the LabeledNarrative schema with fields:'\\
        '- label: Either 'approved' or 'refine''\\
        '- explanation: A short explanation for the decision.'\\
        f'Context:{context}'\\
        f'Narrative:{narrative}'\\
    )\\

\newpage
\section{Evaluation App} \label{app2}
Screenshot of the Evaluation page in the corresponding shiny app

\begin{figure}[htbp]
\includegraphics[scale=0.9]{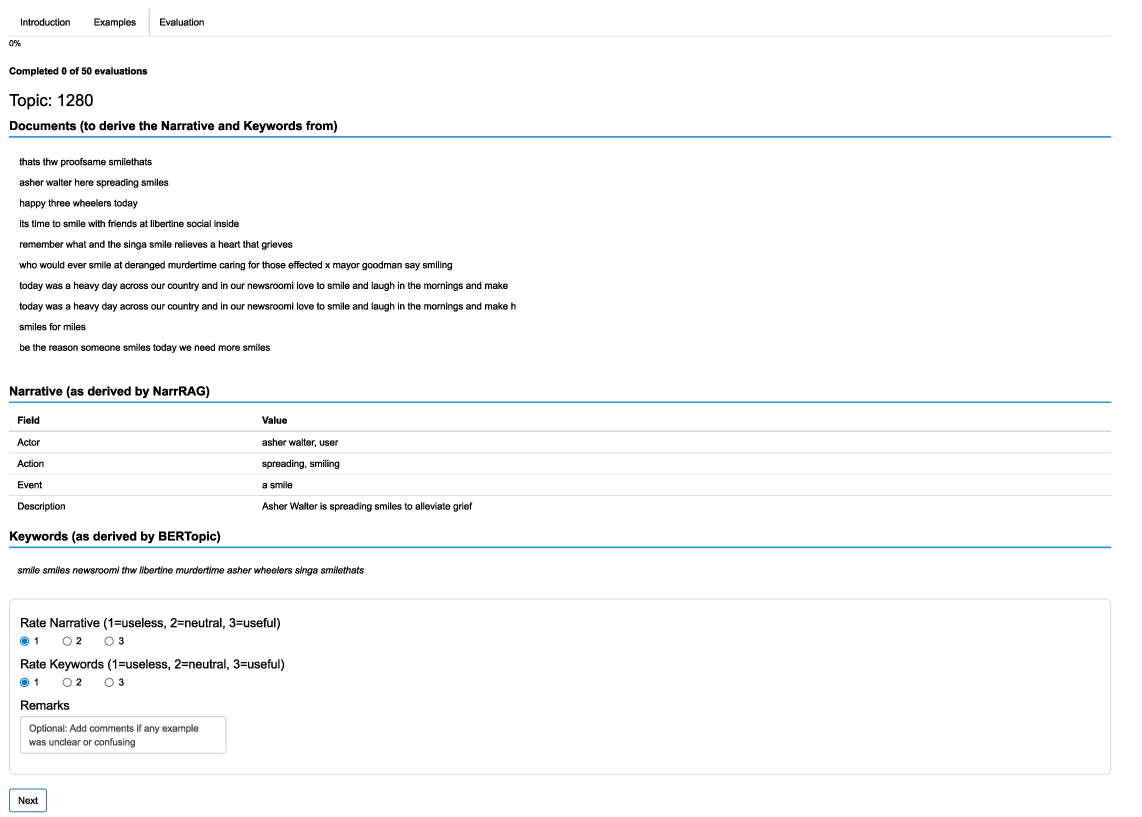}
\caption{Screenshot of the Evaluation App, Evaluation page.}
\label{fig:evaluation_app}
\end{figure}

\newpage

\bibliography{references}

\end{document}